\documentclass[sigconf]{acmart}

\AtBeginDocument{%
  \providecommand\BibTeX{{%
    \normalfont B\kern-0.5em{\scshape i\kern-0.25em b}\kern-0.8em\TeX}}}

\PassOptionsToPackage{numbers, sort, compress}{natbib}

\copyrightyear{2022}
\acmYear{2022}
\setcopyright{acmlicensed}\acmConference[CSLAW '22]{Proceedings of the 2022 Symposium on Computer Science and Law}{November 1--2, 2022}{Washington, DC, USA}
\acmBooktitle{Proceedings of the 2022 Symposium on Computer Science and Law (CSLAW '22), November 1--2, 2022, Washington, DC, USA}
\acmPrice{15.00}
\acmDOI{10.1145/3511265.3550446}
\acmISBN{978-1-4503-9234-1/22/11}

\begin{document}


\title{Non-Determinism and the Lawlessness \\of Machine Learning Code}
\author{A. Feder Cooper}
\authornote{This paper has been updated to align with edits made in 2024 for the first author's Ph.D. dissertation. 
This version, not the version published in the ACM Library, should be considered authoritative.}
\affiliation{%
   \institution{Cornell University}
   \city{Ithaca}
   \state{NY}
   \country{USA}}
   \email{afc78@cornell.edu}

\author{Jonathan Frankle}
\affiliation{%
  \institution{MIT}
  \city{Cambridge}
  \state{MA}
  \country{USA}
}
\affiliation{%
  \institution{MosaicML}
  \city{San Francisco}
  \state{CA}
  \country{USA}
}

\email{jfrankle@mit.edu}

\author{Christopher De Sa}
\affiliation{%
  \institution{Cornell University}
  \city{Ithaca}
  \state{NY}
  \country{USA}}
\email{cmd353@cornell.edu}

\renewcommand{\shortauthors}{A. Feder Cooper, Jonathan Frankle, \& Christopher De Sa}
\renewcommand{\shorttitle}{Non-Determinism and the Lawlessness of Machine Learning Code}

\color{black}
\begin{abstract}
Legal literature on machine learning (ML) tends to focus on harms, and thus tends to reason about individual model outcomes and summary error rates. This focus has masked important aspects of ML that are rooted in its reliance on randomness --- namely, \emph{stochasticity} and \emph{non-determinism}. While some recent work has begun to reason about the relationship between stochasticity and arbitrariness in legal contexts, the role of non-determinism more broadly remains unexamined. In this paper, we clarify the overlap and differences between these two concepts, and show that the effects of non-determinism, and consequently its implications for the law, 
become clearer from the perspective of reasoning about ML outputs as \emph{distributions over possible outcomes}. This distributional viewpoint accounts for randomness by emphasizing the \emph{possible} outcomes of ML. Importantly, this type of reasoning is not exclusive with current legal reasoning; it complements (and in fact can strengthen) analyses concerning individual, concrete outcomes for specific automated decisions. By illuminating the important role of non-determinism, we demonstrate that ML code falls outside of the cyberlaw frame of treating ``code as law,'' as this frame assumes that code is deterministic. We conclude with a brief discussion of what work ML can do to constrain the potentially harm-inducing effects of non-determinism, and we indicate where the law must do work to bridge the gap between its current individual-outcome focus and the distributional approach that we recommend.
\end{abstract}

\ccsdesc[500]{Social and professional topics~Computing / technology policy}
\ccsdesc[500]{Applied computing~Law}


\keywords{machine learning, non-determinism, stochasticity, arbitrariness}

\maketitle

\section{Introduction}\label{sec:intro}

Legal decision logic bears some resemblance with the logic of mathematical functions in that both involve procedures for mapping inputs to outputs. 
When adjudicating a particular case, a magistrate assembles the available evidence, which they supply as parameters to legal rules to inform decisions. 
Just as with mathematical functions, there can be variations in input parameters, which correspond to variations in outcomes.\footnote{For more general background on how legal rules function despite variation in their application, we refer the reader to Fuller~\cite{fuller1965law} and Tamanaha~\cite{tamanaha2004law}.}
Kolber~\cite{kolber2014smoothbumpy} takes this functional analogy a step further, classifying the correspondences between legal inputs and outputs into ``smooth'' and ``bumpy'' types. 
A smooth relationship is one for which gradual changes in inputs map to gradual changes in outputs. Bumpy relationships, in contrast, exhibit discontinuities: 
slight variations in inputs can map to large variations in outputs.\footnote{For example, it may be reasonable to contend that tort law should be smooth, with the amount of harm caused exhibiting a direct and continuous relationship with the degree of compensation owed. 
    However, in practice, tort law is often bumpy: defendants are either liable to provide full compensation (regardless of the particular degree of contributing to harm), or they are not liable at all~\cite[p. 673]{kolber2014smoothbumpy}.} 
Machine learning (ML) --- a discipline within the mathematical tradition --- unsurprisingly seems to follow a similar logic. 
Classification problems resemble Kolber~\cite{kolber2014smoothbumpy}'s concept of bumpiness; varied, continuous inputs become discretized outputs. 
Determining loan-worthiness, for example, is bumpy because a classification model maps personal data to a binary outcome in the set $\{\texttt{grant\_loan}, \texttt{reject\_loan}\}$, typically based on some underlying notion of whether the individual under consideration is likely to repay or default.

This comparison between the work of law and that of ML, in which both are reasoned about as functions, is deceptively attractive. 
At first glance, it seems to mirror the decades-long literature in cyberlaw that has considered the law and if/then code rules\footnote{Either as a type of architecture~\cite{lessig1999horse, lessig2009code} or a modality on its own~\cite{grimmelmann2005reg}.} 
to be complementary modalities that regulate and mediate human experience~\cite{lessig1999horse, grimmelmann2005reg, bamberger2010risk, citron2008dueprocess,  lessig2009code, reidenberg1997code}. 
It is thus perhaps intuitive to consider stretching this analogy further: 
to treat the mathematical-functional similarity of the law and ML as a rationale for christening ML as the latest type of code-imbued regulator. 
To stretch this even further, if ML can be fashioned to design new ``microdirectives'' or usher in a new era of ``personalized law,'' as some legal scholars contend~\cite{casey2015rulesandstandards, fagan2019discretion}, then perhaps ML could breathe new life into the succinct cyberlaw refrain that ``code is law''~\cite{reidenberg1997code, lessig2009code}. 
That is, rather than using this widely-quoted shorthand to stand in for the more-precise (but still abbreviated) ``code is constitutive of law''~\cite[p. 675]{bamberger2010risk}, ML code could literally be used to generate law.

And yet, while it might be appealing to take these steps to connect the nascent field of ML law with its older cyberlaw sibling, upon deeper examination the comparison between ML and the law via functions does not hold up. 
For one, as much legal scholarship acknowledges, the mechanism by which ML translates from inputs to outputs fundamentally differs from analogous mechanisms in the law~\cite{mulligan2018governance, mulligan2019ml, citron2014scored, kroll2017accountable, hausman2021rigged, barocas2016data, citron2022privacy}. 
The law has a variety of mechanisms --- rules, standards, factors tests, etc. --- each accompanied with justifications for (and amendments regarding) their use, as well as a long record in jurisprudence of their application to specific cases. 
In contrast, ML may behave like a function, but we often do not understand how that function works. 
In ML systems, we can have full access to both the inputs and subsequent outputs, while having no clear understanding of \emph{how} the mapping from one to the other occurred. 
In other words, unlike the law, ML functions defy explanation and reasonable justification, which in turn raises fundamental questions about the legitimacy of using ML as a decision-making tool and muddies the ability to determine accountability when these tools cause harms~\cite{cooper2022accountability, creel2022leviathan}.\footnote{Clarity of explanation in legal contexts, however, is not a given. 
    As Fuller~\cite{fuller1965law} notes, ``It is easy to assert that the legislator has a moral duty to make his laws clear and understandable. But this remains at best an exhortation unless we are prepared to define the degree of clarity he must attain in order to discharge his duty. The notion of subjecting clarity to quantitative measure presents obvious difficulties. We may content ourselves, of course, by saying that the legislator has at least a moral duty to try to be clear. But this only postpones the difficulty, for in some situations nothing can be more baffling than to attempt to measure how vigorously a man intended to do that which he has failed to do. ... [However,] good intentions are of little avail. ... All of this adds up to the conclusion that the inner morality of law is condemned to remain largely a morality of aspiration and not of duty. Its primary appeal must be to a sense of trusteeship and to the pride of the craftsman''~\cite[pp. 42-43]{fuller1965law}. 
    It is reasonable to argue, though out of scope for this paper, that ML does not have an analogous ``sense of trusteeship'' on which the public can rely.} 
In short, ML's problem with \textit{explainability} shows how the analogy essentially and inescapably falls short; both the law and ML may behave like functions, but functions that are fundamentally different in kind.

This analogy falls short in another fundamental way --- one that is significant enough for us to pause attempting to close the loop between cyberlaw, code-is-law scholarship and legal scholarship about ML, but has thus-far remained under-explored. 
Code that follows if/then logic --- the type of code addressed in cyberlaw literature~\cite{lessig1999horse, grimmelmann2005reg, bamberger2010risk, citron2008dueprocess} --- is \emph{deterministic}: 
it specifies behaviors to execute (the ``then'') when certain, specified conditions (the ``if'') are met. 
Importantly, ML code does not execute if/then rules. 
Instead, the ML training process is random in nature; it exhibits \emph{stochasticity} and \emph{non-determinism}.\footnote{Non-determinism and stochasticity are not unique to ML, but rather are features of many types of randomized programs (including programs and protocols that predate the Internet and cyberlaw). 
    Nevertheless, the advent of ML applications in public life, and the social valences these applications carry, has brought urgency to clarifying these concepts in relation to ML.} 
We explore the meaning of these terms in detail later in this paper (Section~\ref{sec:nondeterminism:ml}). 
For now, it suffices to provide an intuition: deterministic code ensures that computing with the same inputs yields the same outputs; stochasticity and non-determinism, in contrast, can cause two similar training procedures to produce vastly different results in practice~\cite{forde2021disparate, qian2021variance, cooper2024variance}. 

In the remainder of this chapter, we explain how stochasticity and non-determinism play a fundamental role in the behavior of ML systems. 
While some legal scholarship has begun to reason about the relationship between stochasticity and arbitrariness~\cite{creel2022leviathan, bambauer2022diff}, the role of non-determinism more generally remains unexamined. 
We argue that a more precise understanding of non-determinism is essential for reasoning about questions concerning the regulability, legitimacy, and accountability of ML decision-making tools. 

Our first contribution is to show that the emphasis on individual errors and error rates in existing legal scholarship is concealing other important issues in ML that are rooted in non-determinism. 
While focusing on individual outcomes and error rates for specific models is important --- and intuitive, given that it parallels case-based analysis in the law --- it nonetheless provides a limited view of behavior of ML. 
We clarify the distinction between stochasticity and non-determinism more broadly construed, and show that the effects of non-determinism, and consequently its implications for the law, instead become clearer from the perspective of reasoning about ML outputs as \emph{distributions or patterns over possible outcomes}. 
The key difference is that this viewpoint accounts for randomness and other types of non-determinism by providing a window into the \emph{possible} outcomes of ML. 
Importantly, this type of reasoning is not exclusive with current legal reasoning; it complements (and in fact can strengthen) analyses of individual, concrete outcomes for specific automated decisions (Section~\ref{sec:nondeterminism:ml}). 

By illuminating the important role and potential effects of non-determinism, we then demonstrate that ML code falls outside of the cyberlaw frame, which assumes deterministic code (Section~\ref{sec:nondeterminism:law}).
Even if this frame can be expanded to include the stochastic elements of ML, we discuss how it cannot be extended to non-de\-ter\-min\-is\-tic elements more generally.  
Lastly, we conclude with a brief discussion of what work ML can do to constrain the potentially harm-inducing effects of non-determinism, and we indicate where the law must do work to bridge the gap between its current case-based analysis of ML systems and the pattern/distributional analysis that we recommend (Section~\ref{sec:nondeterminism:conclusion}).

\section{Non-determinism and Stochasticity}\label{sec:nondeterminism:ml}

Legal literature regarding the empirical performance of ML tools tends to focus on issues of accuracy~\cite{lehr2017legalml}~\cite[pp. 1249-50]{brennanmarquez2019plausiblecause}~\cite[pp. 9,12]{calo2021modeling}~\cite[p. 1253]{citron2008dueprocess}.\footnote{Work on fairness typically focuses on accuracy, as well, by emphasizing differences in inaccuracy via error rates, and the resulting disparate impact, for protected demographic groups.} 
This work typically evaluates ML in terms of individual decision outcomes in relation to the harms these outcomes cause, and uses summary error rates to draw conclusions about a particular model's accuracy. 
Solely focusing on the accuracy of specific inference outcomes and summary rates can conceal other important issues implicated by non-determinism, which are also important factors to consider in legal analyses of ML technology. 
To make this case, we first must establish definitions for non-determinism and stochasticity, as there are nuanced differences and overlap between the two terms.

\begin{definition}
\label{def:nondeterminism}
\textbf{Non-determinism} is a property of processes for which supplying the same inputs can produce different outputs.
\end{definition}

As a result, non-deterministic outcomes are uncertain. 
This is in contrast to deterministic if/then logic, for which the same inputs produce the same outputs. 
Stochasticity also satisfies Definition~\ref{def:nondeterminism}; however, it places additional conditions on the form that uncertainty can take.

\begin{definition}
\label{def:stochasticity}
\textbf{Stochasticity} is a property of non-deterministic processes whose outcomes can be reasoned about using probability theory.
\end{definition}

In other words, the non-determinism of stochasticity specifically comes from randomization that can be understood using probability. 
Following these definitions, we can think of stochastic decision-making processes as non-deterministic; 
however, non-deterministic decision-making processes are not necessarily stochastic, since they cannot always be reasoned about using the laws of probability. 

Machine learning is grounded in probability and statistics, and thus is fundamentally stochastic in nature. 
In practice, however, it is also common for ML to exhibit non-determinism beyond this stochasticity. 
While the formal specification for an algorithm is stochastic, its implementation and execution in software and hardware can introduce non-determinism that is not stochastic. 
We can attempt to apply the rules of probability to reason about this behavior, but we are not guaranteed that our conclusions will be sound. 
A notable example of this non-determinism comes from the popular PyTorch library.\footnote{We refer to \textcolor{blue}{\href{https://pytorch.org/docs/stable/notes/randomness.html}{PyTorch}} for discussion about limiting the sources of software and hardware non-determinism in ML training pipelines. 
    At the time of writing, PyTorch offers a ``deterministic mode'' that, at the cost of significant run-time slowdowns that may not be feasible for all application developers, enforce determinism in software operations (where possible).} 
When prepared for execution on a computer at training time, PyTorch makes dynamic choices regarding how to run the code, which optimize for run-time speed and, in doing so, introduce non-stochastic non-determinism to the learning process.

\subsection{Related Work: ML Stochasticity and the Law}\label{sec:nondeterminism:prior}

The legal literature that discusses uncertainty and subsequent impressions of arbitrariness in ML decision-making does not reckon with this practical reality.
Rather, in talking about algorithms, and more specifically their error rates or individual outcomes, this literature regards ML in stochastic terms. 
For example, Bambauer et al.~\cite{bambauer2022diff} coins the term ``Small Change Makes a Big Difference'' (SCMBD) to analyze the risks to due process that can come from disproportionate outcomes on similar inputs due to the stochastic nature of ML training pipelines~\cite[pp. 2378-2383, pp. 2396-2397]{bambauer2022diff}. 
Their discussion makes no mention of how other sources of non-determinism further expand this category of risk.

In another recent example, Creel and Hellman~\cite{creel2022leviathan} take a formal philosophical approach to understanding what is precisely connoted by  criticisms of ``arbitrariness'' in ML system outputs. 
They break down their analysis of what is arbitrary in three different respects: ``unpredictable,'' ``unconstrained,'' and ``unreasonable'' behaviors of these systems.\footnote{In their discussion of ``arbitrary'' as ``unconstrained,'' Creel and Hellman~\cite[p. 3-4]{creel2022leviathan} call algorithms ``rule-based'' in close proximity to discussing legal rules and standards. 
    We do not believe that stochastic algorithms are ``rule-based'' in the same sense as legal rules; 
    however, discussing this distinction is out of scope for this paper.
} 
In their discussion, they claim that arbitrariness of ML systems in itself is not the problem; rather, the problem is ``the systematicity of their arbitrariness'' that may ``irrationally [exclude] a person from a significant number of important opportunities''~\cite[p. 2]{creel2022leviathan}.\footnote{douek~\cite{douek2021moderation} makes a related but different point about shifting legal understanding away from individual outcomes. She call for a shift ``from an individualistic approach to a probabilistic one''~\cite[p. 789]{douek2021moderation}. 
    douek makes an important intervention regarding the inevitability of error in ML applications, particularly at scale, but ultimately focuses on individual model error rates and makes an argument predicated on the ability to reason about probabilities, and thus is not examining the same concepts with which we concern ourselves here.}  
In relation to this claim, they add ``To the extent that an algorithm governs the decision, it will produce the same result when run on the same inputs. If the \textbf{algorithm} contains a degree of \textbf{randomness} within it, ... it is still \textbf{reproducible} at a higher level of abstraction''~\cite[pp. 3-4, emphasis added]{creel2022leviathan}. That is, they describe a \emph{model} demonstrating deterministic behavior. A particular model produced from an algorithmic learning procedure is deterministic --- always producing the same output given the same input (Section~\ref{sec:nondeterminism:example1}); however, as we have discussed above, the entire procedure that produces such a model is \emph{not} deterministic. 

Put differently, implicit in the reasoning in Creel and Hellman~\cite{creel2022leviathan} is that the uncertainty at play in ML can be reasoned about using probability. 
It is probability theory that enables the systematic, ``higher level of abstraction'' of reasoning about the overall, expected behavior of stochastic \emph{algorithms}, and whether those behaviors are systematically, arbitrarily unfair (according to a particular fairness criterion). However, in contrast to abstract algorithm specifications, the implemented, run-time behavior of ML \textit{pipelines} and \textit{systems} introduces non-stochastic non-determinism --- non-determinism that is \emph{not} systematic, in the sense that it cannot be reasoned about analytically with the guarantees of probability theory. This is not a distinction without import; in contrast to Creel and Hellman's claim about reproducibility in relation to what we understand as stochastic-related arbitrariness, this kind of non-determinism is a well-known contributor to the reproducibility crisis in ML~\cite{raff2019reproducibility, bouthillier2019reproducibility}. 
Non-stochastic non-determinism thus suggests a different kind of arbitrariness from that discussed in Creel and Hellman~\cite{creel2022leviathan}, and it, too, can have significant impacts on normative concerns like fairness~\cite{qian2021variance} (Section~\ref{sec:nondeterminism:example2}). \\

\noindent In short, though prior legal literature on ML and arbitrariness sometimes engages with elements of stochasticity, it does not account for the role of other forms of non-determinism. 
In the remainder of this section, we 
explain via simple synthetic examples how the presence of non-determinism calls into question essential assumptions about the fundamental nature of accuracy in ML. Moving away from analyses of individual outcomes to thinking about \emph{distributions/patterns over possible outcomes} can expand legal scholars' understanding of the behavior of ML tools. 
In particular, reasoning about \textit{probability} distributions over possible outcomes is useful for understanding the impacts of stochasticity (Section~\ref{sec:nondeterminism:example1}). 
While probability is not similarly useful for analytically reasoning about other sources of non-determinism, our approach can still highlight empirically the importance of the role of non-determinism in ML and the potential harms it can cause (Section~\ref{sec:nondeterminism:example2}). 

\subsection{Distributions over Individual Outcomes}\label{sec:nondeterminism:example1}

\begin{figure}[t!]
  \begin{center}
    \includegraphics[width=0.45\textwidth]{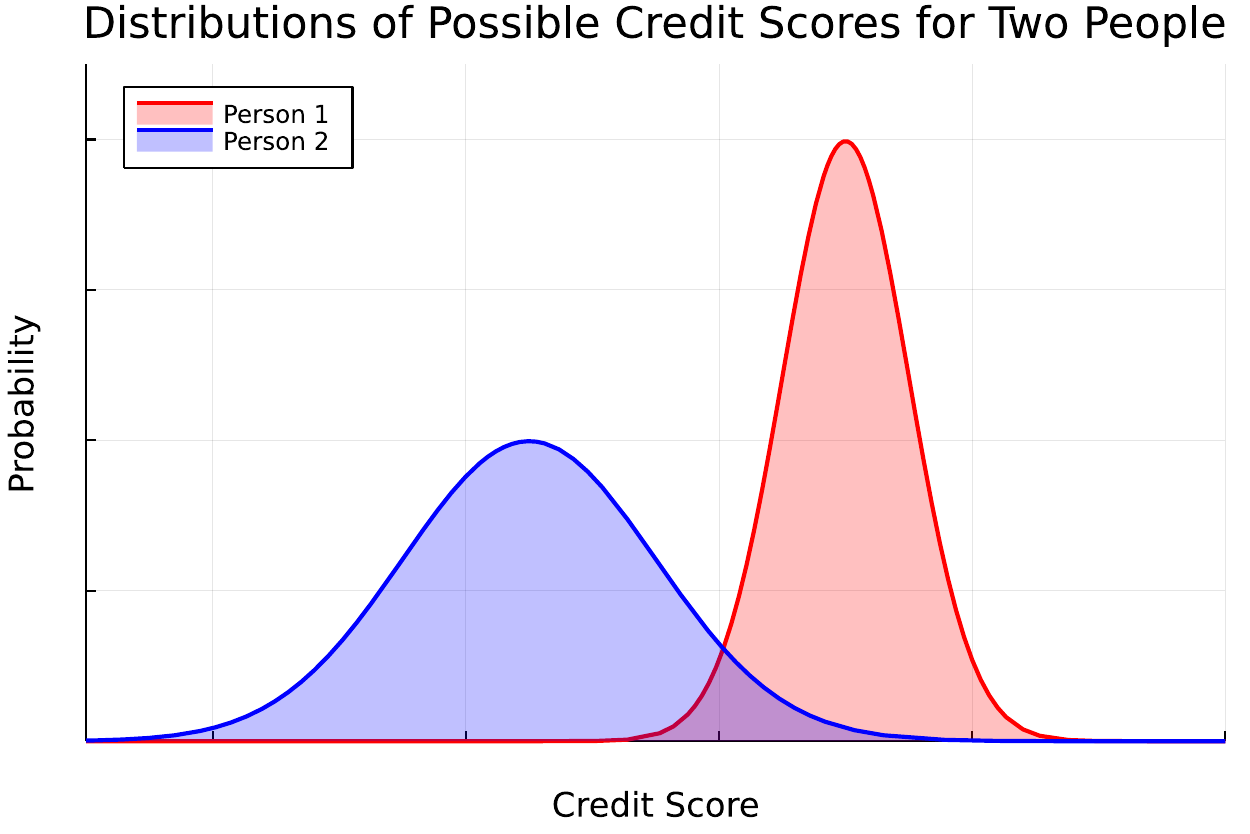}
    \caption{Synthetic probability distributions for possible predicted credit scores of two different individuals.}
	\label{fig:loans}
  \end{center} 
\end{figure}

We first consider a synthetic ML system that aims to determine individuals' creditworthiness by predicting their credit scores. 
The developers write a snippet of code to achieve this task --- a procedure for training models to predict individuals' credit scores. 
The execution of this code to actually train a model exhibits stochasticity: running this one piece of code multiple times on different subsets of the training data will result in multiple trained models that vary in comparison to one another. 
If we were to take many such models and supply them with the same individual as input, the corresponding outputs would yield a distribution over possible credit score outcomes for that individual. 
We illustrate this in Figure~\ref{fig:loans} for two individuals. 
In other words, since this process yields a distribution over possible credit scores for each individual --- and not just a single credit score --- predicting an individual's credit score is not a deterministic function of the code written by the engineer to train ML models. 
Rather, credit score for an individual is a function of the procedure that this code can execute; 
it is a function of executing model training, which exhibits stochasticity (as a function of the specific training data examples used for training) and thus a distribution of possible outcomes for different individuals.

The viewpoint of distributions over possible outcomes shown in Figure~\ref{fig:loans} illustrates a problem: 
the two individuals have overlapping credit score distributions  (shown in purple). This means that it is possible that there is some subset of models, produced by the stochastic training process, for which we cannot distinguish between these two individuals in terms of their credit scores. 
And yet, in looking at each of their distributions overall, there are all clearly cases where they do not overlap and are thus clearly distinguishable. 
That is, from its distributional perspective, this figure shows that it is possible to produce models that suggest contradictory results: 
some models are able to distinguish these individuals via different credit scores, while it is possible that some models are \emph{not} able to discern a difference. 
Instead of all models from this training process having the ability to clearly distinguish between or to equate these two individuals via credit scores, both contradictory possibilities are suggested by this distributional viewpoint. 

This ambiguity complicates what accuracy means for a model produced by this training process, because it is not clear what a ``correct'' model should do with respect to how it views these two individuals. 
Is it ``correct'' to model them as distinguishable, or ``correct'' to model them as indistinguishable, in terms of their credit scores? 
It is impossible to say with 100\% certainty, since there is no notion of ground truth credit score.\footnote{This is in contrast to applications for which we can reasonably say that there is a ground truth, such as a computer vision system that distinguishes between cats and dogs; an example input is either a cat or a dog, not both.} 
Put differently, this figure indicates that there is a meta-problem of not being able to draw a firm line between correctness and incorrectness for models trained by this process. 
This issue of being unable to draw a clear boundary between correctness and incorrectness illustrates how model output decisions can exhibit non-determinism: 
for the different inputs, depending on the model, the outputs for those inputs may be distinguishable or may be indistinguishable. 

So far, we have limited our discussion of non-determinism to stochasticity (Definition~\ref{def:stochasticity}) --- in particular, the stochasticity resulting from training models on different subsets of the training data or from different examples drawn from the same data distribution. 
In practice, the other sources of non-determinism that we describe above can contribute to the results described in Figure~\ref{fig:loans}.
Moreover, it may not be immediately clear how each source contributes to the outcome predictions and impacts their associated probabilities. 
In other words, the distributional approach in Figure~\ref{fig:loans} clarifies that the predictions can fluctuate, but it conceals how stochasticity and other sources of non-determinism interact to produce those fluctuations --- a point to which we return in Section~\ref{sec:nondeterminism:law}, where we discuss the regulability of ML code.

For now, we observe that the legal literature discussed in Section~\ref{sec:nondeterminism:prior} touches on the stochasticity that contributes to examples like this one, but it does so in a manner different from the distributional picture we show here. 
Bambauer et al.~\cite{bambauer2022diff} discusses how stochasticity can cause \textit{a particular model} to exhibit SCMBDs that affect due process. 
Similarly, Creel and Hellman~\cite{creel2022leviathan} discuss how \textit{a particular model} exhibits deterministic outputs; their concern is that, at the scale of multiple decisions across multiple models for different tasks, there may be a pattern of arbitrary discrimination against certain individuals. 
In relation to Figure~\ref{fig:loans}, these works engage with stochasticity at the point in which there is one model producing a concrete credit score for each individual, rather than the distribution of possible model outputs for these individuals. 
It is only in the setting they rely on --- after we have selected a particular model to use for predicting credit scores --- that we can think about deterministic outputs. 
That is, by picking a particular model that encodes a specific function, we have locked in a deterministic score for each individual. 
We can then move from reasoning about distributions over possible outcomes of credit scores for individuals, as indicated in Figure~\ref{fig:loans}, to thinking about deterministic, concrete outcomes, which are conditional on the model we have chosen. 

Given one specific model, with deterministic outcomes for each individual input, it becomes possible to perform analyses concerning the inaccuracy of individual outcomes, associated harms, and metrics like error rates to capture summary information about a model's overall performance across a sample of inputs, as Bambauer et al.~\cite{bambauer2022diff} and Creel and Hellman~\cite{creel2022leviathan} both do. 
But, importantly, at the distributional level conveyed in Figure~\ref{fig:loans}, concepts like accuracy remain slippery. 
In reasoning about possible rather than specific model outcomes, this level of abstraction makes the potential areas of uncertainty in trained models --- whether due to stochasticity or other sources of non-determinism --- more transparent. 
It clarifies how the possibility of different outcomes can have the effect of muddling the distinction between correctness and incorrectness, and opens up the possibility of trying to untangle sources of non-determinism and their respective normative considerations regarding arbitrariness, which we discuss further in Section~\ref{sec:nondeterminism:law}.

\subsection{Patterns over Models}\label{sec:nondeterminism:example2}

\begin{figure}[t!]
  \begin{center}
    \includegraphics[width=0.45\textwidth]{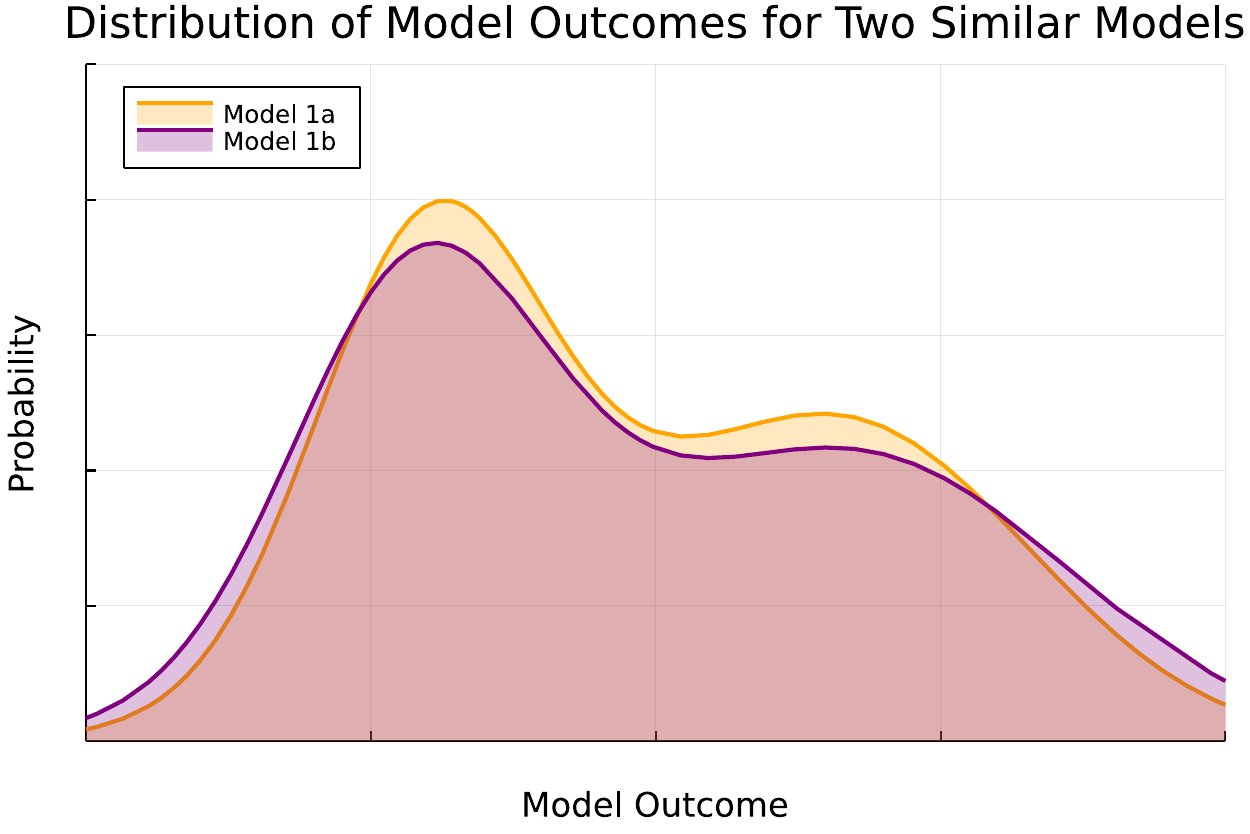}
    \caption{Synthetic patterns of model outcomes for two models trained on the same training data for the same task, using the same algorithm and data, but possibly different computers with different hardware random seeds. Non-determinism in the training process yields different patterns of model outcomes.
    We visualize this pattern like a probability distribution, but there are no guarantees that we can reason reliably about this pattern with the tools of probability.}
	\label{fig:models}
  \end{center}
\end{figure}

Reasoning over distributions of outcomes does not just apply to thinking about how outcomes for fixed individual inputs may vary based on choice of model. 
This view can also help reason about how non-determinism affects models trained from the same stochastic training process. 
Figure~\ref{fig:models} shows patterns\footnote{In the camera-ready version of this paper, we used the word ``distributions'' to describe this effect as well, since we could not think of a better term to use at the time. 
    However, this was a poor choice on our part, since the type of non-determinism we describe in this section cannot be reasoned about with probability, and ``distribution'' most typically implies that we are talking about a ``probability distribution.''
    Joan Feigenbaum suggested we use the word ``pattern'' instead, so we make that change here. 
} over model outcomes for two models trained using the exact same procedure and, unlike the prior example, the models are trained using the same software random seed, which functions to supply the algorithm with the exact same training data. 
With this setup, we have controlled for every possible source of stochastic non-determinism in the training process. By using the same random seed, we should be able to consistently reproduce the same deterministic model, aligning with Creel and Hellman's conception of the training process (Section~\ref{sec:nondeterminism:prior}), and thus the two curves in Figure~\ref{fig:models} should completely overlap.\looseness=-1 

The reason they do not overlap is because of non-stochastic non-determinism that affects their respective training processes differently. 
For example, differences in hardware random seeds, which we cannot control in software code, cause the training process to produce different models that reflect different underlying deterministic functions. 
Ideally, even with non-determinism in ML software packages and across hardware, repeated runs of similar or identical training processes would result in outcome distributions that are reasonably similar to each other (as, one could argue, is the case in Figure~\ref{fig:models}, since the curves roughly overlap). 
If the models' patterns of outcomes do not vary too much, then at least we can be confident (however informally) that picking any of them as the specific model to deploy is a reasonable choice, as each model indicates performance roughly representative of all the models that were trained. 
In other words, it might be fine to avoid the issue of drawing a line between which models are correct and which are incorrect, because all of the models are effectively the same. 

Of course, though, the models are not \textit{exactly} the same, which may have significant consequences at the more granular level of individual outcomes. 
They may differ in a way that is not semantically meaningful, or may exhibit more uncertainty in connection with a protected attribute value, such as a particular gender or race. 
Importantly, this type of uncertainty, not being stochastic in nature, is not amenable to reasoning with the tools of probability; it is not amenable to the same reasoning about arbitrariness and individual outcomes in Creel and Hellman~\cite{creel2022leviathan}, which is implied to be predicated on uncertainty due to stochasticity.

More generally, the fact that models are not completely identical requires us to shift our thinking about ML. 
This pattern-level view clarifies that we should be thinking of one run of an ML training process as learning \emph{a} possible pattern over possible outcomes, rather than \emph{the} singularly correct pattern over possible outcomes. 
Additionally, unlike the synthetic example in Figure~\ref{fig:models}, in practice it is common for model outcome patterns to vary more significantly due to non-determinism~\cite{bouthillier2019reproducibility, raff2019reproducibility, sivaprasad2020hpo, cooper2021hpo, forde2021disparate, qian2021variance}. 
In such cases, it will not necessarily be clear if there is a representative model in the group --- if there is a model that is more ``correct'' than the others. 
Once again, due to non-determinism, drawing a firm boundary between correctness and incorrectness is ill-defined. 
As with the example in Figure~\ref{fig:loans}, this example similarly raises questions of how to legitimately pick a model that we can be confident will yield robust and reliable performance.\footnote{We could also extend the first example be plotted at the model level, rather than individual level, with outcomes on the $x$-axis and then the probability on the $y$-axis; 
    in this case, where we only look at stochasticity as a function of the training dataset, we would have a picture that looks perhaps a lot like Figure~\ref{fig:models}, but the terminology would change to reflect that we could reason about this plot as containing probability distributions.
}

This question is not just of theoretical relevance. 
In practice, non-determinism can cause resulting model outcome distributions to vary so much that, for a particular input, models can yield wildly inconsistent results. 
To make this more concrete, we describe an example in the ML literature that demonstrates the effects of such non-determinism. In recent work, Forde et al.~~\cite{forde2021disparate} and Qian et al.~\cite{qian2021variance} investigated how the impact of stochasticity\footnote{And we performed the largest such study on stochasticity in fairness contexts in other work that followed~\cite{cooper2024variance}.} 
and non-determinism on training models using similar training procedures can impact model fairness. Qian et al.~\cite{qian2021variance} published an extensive empirical study, in which they repeatedly trained models with identical training procedures, using the same software random seed and thus exactly the same training data examples across runs. 
In theory, this setup should control for stochasticity in different model outputs; by using the same training data and same training procedure, the models produced should be the same. 
However, the realities of running ML code in practice differ from what we expect in theory. Qian et al.~\cite{qian2021variance} makes the stakes of this point unimpeachably clear by comparing fluctuations in the resulting model outcome distributions. 
In particular, they computed common algorithmic fairness metrics to probe how fairness measurements varied for these (theoretically identical) models, and found that fairness measurements could vary by up to 12.6\%.
This degree of variance was so significant that, in some cases, it was possible for one trained model to pass US legal compliance rules regarding fairness thresholds on the test set, while another model could violate those same requirements~\cite[p. 2]{qian2021variance}.

In other words, Qian et al.~\cite{qian2021variance} illustrates clearly how non-stochastic non-determinism can have a significant impact on fairness in the distribution of possible modeling outcomes. 
This result indicates that picking any one specific model to deploy --- which then could be examined in terms of individual errors and error rates, fairness-related or otherwise --- is a non-trivial task. 
Non-determinism necessarily has an unpredictable role in the specific outcomes of training models, as evidenced by the resulting evaluation of test error to understand generalization. 
When this unpredictability leads to wide variability in metrics like fairness, this then raises fundamental questions not just about the fairness of particular models, but about the fairness of the process by which those models were trained. 
We may try to the best of our ability to control for models to be trained in the same way, and yet they may still exhibit vastly different fairness levels. 
Given this non-determinism, how can we be sure, especially when training just a few models under limited computational resource budgets, that the model we have selected to deploy in practice is representative of what is (at least close to) maximally possible in terms of fairness?

Questions like these, let alone their answers, are not clear from looking at individual outcomes or error rates for single models alone. 
Instead, it is looking at patterns and distributions over outcomes that raises questions about the legitimacy model-producing processes, through indicating how the resulting models from those processes can fluctuate in important ways. 
This distributional/pattern-level view provides information that can help us interrogate whether the process for training ML models for a specific task is robust enough to justify the use of \textit{any} such model produced from that process. 
By robust we mean that, even in the presence of non-determinism, the resulting variation in the behavior of possible ML models --- whether variation in model outcome distributions, or variation in outcomes across models for particular individual inputs --- is not the product of happenstance, for example resulting from a particular hardware-software interface implementation. 

The example of Qian et al.~\cite{qian2021variance} arguably does not meet this definition of robustness, given the large variance in fairness metrics across the distribution of models they produced.\footnote{Neither do the individual models trained in Cooper et al.~\cite{cooper2024variance}; 
    however, the ensemble models trained in that work are more robust in this sense.} 
This becomes especially clear when one considers how such variance in fairness could impact due process~\cite{lehr2017legalml}~\cite[pp. 1249-50]{brennanmarquez2019plausiblecause}~\cite[pp. 9,12]{calo2021modeling}~\cite[p. 1253]{citron2008dueprocess} --- 
if a particular chosen model by chance demonstrates poor performance with respect to fairness, in turn leading to a greater number of unfair individual outcomes in practice. 

\section{Non-deterministic Code Is Lawless}\label{sec:nondeterminism:law}

In moving from looking at individual errors and model error rates to reasoning about distributions and patterns of outcomes, we have seen how the non-determinism inherent in ML can raise key questions concerning the legitimacy of using ML-driven processes in decision-making. 
We have seen, too, how non-determinism can directly effect harms at the individual level, in cases in which a training process is not sufficiently robust to guarantee that its resulting models behave similarly for key metrics, such as fairness. 
In short, our discussion thus-far has indicated that non-determinism can have significant, detrimental effects on the behavior of ML code. 
While there are different types of non-determinism, we have shown that prior work in legal ML focuses on non-determinism that is stochastic (Definition~\ref{def:stochasticity}).\footnote{However, upon further reflection, we realize that this work has not studied this sufficiently; we defer additional study to future work.} 
While this type of non-de\-ter\-min\-ism is amenable to analysis using probability, other types of non-determinism in ML, such as the specifics of the hardware platform used to execute training processes, do not follow the same logic (Definition~\ref{def:nondeterminism}). 
As a result, work that has engaged with arbitrariness of ML decisions purely in stochastic terms has missed this crucial aspect of non-determinism and its relationship to arbitrariness. 

One of the important consequences of this omission has to do with an implied, uncomfortable relationship between arbitrariness and necessity in ML. 
As we briefly discussed in the introduction, the stochasticity of ML is one of its core strengths that separates it from non-stochastic decision systems; 
it is the property that makes it possible for ML to model phenomena that are too complex to specify exhaustively using if/then deterministic rules~\cite{murphy2022pml1}. 
Yet, stochasticity can also produce variable outcomes for the same inputs, and these variations can suggest contradictions that call the reliability of ML into question (Section~\ref{sec:nondeterminism:example1}). 
Moreover, these potential contradictions are less intuitive to grasp than the outputs of deterministic decision processes. At times, they might even seem like software bugs, rather than an artifact of a necessary feature of ML,\footnote{For work on the elusive boundary between bugs and inherent features in ML, please refer to Cooper et al.~\cite{cooper2022accountability}. 
    More generally, delineating what constitutes a bug for randomized programs is a philosophical question, which has long remained unresolved in the Programming Languages research community~\cite{kozen1981semantics, kozen1983pdl}.} 
which itself can further cast doubt on reliability. 
Due to this seeming double bind, it makes sense that legal literature about ML has tried to parse the cases in which the stochasticity-induced arbitrariness present concerns for the law.

However, other sources of non-determinism do not exhibit the same conflict. 
Lack of expressivity in hardware-software interfaces, inability to control hardware random seeds, and missing APIs for fine-grained control of run-time optimization of ML code all contribute to non-stochastic non-determinism; but, they are not necessary features of ML. 
They are not inherent to machine learning in theory; they are a reality of its practice. 
As a result, this source of non-determinism suggests potential sites for future reliable ML research. 
Nevertheless, in the interim, ML software and hardware ecosystems inject non-determinism into training processes, which affects the patterns of overall outcomes such that they deviate non-probabilistically from what is theoretically expected. 
What makes this especially challenging is that, as we demonstrated in our synthetic examples (Section~\ref{sec:nondeterminism:ml}), it is not always immediately clear which kind of non-determinism is responsible for impacts on the resulting distribution of outcomes, which further complicates our ability to reason about outcomes using the tools of probability.


More generally, taken together, both sources of non-determinism can make it very difficult to reason about the difference between correctness and incorrectness in ML program behaviors, thus making accuracy a fuzzy concept that is difficult to pin down.\footnote{It is also worth noting that the approximate computing concept of the trade-off between accuracy and efficiency~\cite{cooper2021eaamo, cooper2022avs}, and more generally using a temporal lens to analyze outcomes~\cite{susser2022time}, further complicates our understanding of accuracy in ML.} 
And yet, in the existing legal literature on ML, the issue of inaccuracy and accuracy, particularly at the individual model level, has been a dominant theme~\cite{douek2021moderation, douek2022formalism, brennanmarquez2019plausiblecause, citron2014scored, barocas2016data, calo2021modeling, lehr2017legalml}. 
For the law to adequately contend with non-determinism, we have argued that the legal literature must shift to also consider the viewpoint of distributions/patterns over outcomes, as this viewpoint indicates how non-determinism fundamentally problematizes our understanding of accuracy. 

Based on this prior discussion, we now argue that this will also require a shift in the dominant thread of cyberlaw thinking that echoes the refrain that ``code is law.''\footnote{This phrase, which originated from work in  Reidenberg~\cite{reidenberg1997code}, has been further developed and revised~\cite{lessig1999horse, lessig2003cycle, grimmelmann2005reg}, and then ultimately itself codified in Lessig~\cite{lessig2009code}. 
    It has since been partially adapted to account for the new kinds of experiences that ML (particularly robotics) will mediate~\cite{calo2015robotics, balkin2015robotics}.} 
In brief, ``code as law'' stands in for the idea that code does the work of law; code, like the law, is a modality for regulating and mediating human behavior~\cite{lessig1999horse, grimmelmann2005reg}. 
As Grimmelmann~\cite{grimmelmann2005reg} summarizes in more detail, ``code is law'' captures the idea that ``software itself can be effectively regulated by major social institutions, such as businesses or governments. ... If other institutions can regulate software, and software can regulate individual behavior, then software provides these institutions an effective way to shape the conduct of individuals''~\cite[p. 1721]{grimmelmann2005reg}.\footnote{Importantly, this understanding of ``code as law'' grew out of legal scholarship that was reckoning with the advent of the Internet. 
    In particular, this scholarship was concerned with ``decisions about the technical future of the Internet,'' which it considered to be ``important questions of social policy ... [that would] have the force of law even as they def[ied] many of our assumptions about law''~\cite[p. 1721]{grimmelmann2005reg}.}

In the extensive literature that has followed from Lessig~\cite{lessig2009code}'s codification of the concept, various scholars have built on and problematized different aspects of ``code is law''~\cite{grimmelmann2005reg, bamberger2010risk, calo2015robotics}, such that it has ultimately remained a resonant and powerful frame for thinking about technology. 
However, the work that contends with this concept tends to (often implicitly) assume a deterministic view of code. It considers code to be a set of automated if/then rules that ensure consistent decisions --- and can be institutionally regulated to ensure consistent decisions --- as it works to enable and constrain human behavior~\cite[pp. 1721, 1728-1732]{grimmelmann2005reg}~\cite[p. 676]{bamberger2010risk}~\cite[p. 1253]{citron2008dueprocess}. 
In this view, code can concretely specify rule-like (rather than standard-like)\footnote{It is perhaps interesting to consider --- though out of scope in this short paper --- how non-deterministic ML code may more closely resemble standards than rules.} 
relationships between inputs and outputs that are ``free from ambiguity''~\cite[p. 1723]{grimmelmann2005reg}. 
Put simply, this conception of code maps nicely to if/then rules that resemble those in the law. 
Yet, as we have seen throughout this paper, the assumption of deterministic code does not hold for ML: 
due to its statistical nature, ML code does not operate by deterministic if/then rules. Instead, due to non-determinism, it is as if both the ``if'' and the ``then'' are fuzzy; they are not specifiable in concrete terms. 
It is therefore natural to ask: what does non-deterministic code do to an idea of ``code as law'' that is predicated on determinism?

We attempt an answer in a (sort-of) proof by contradiction. 
We begin by assuming that ``code as law'' still holds for the non-deterministic code of ML. 
From there, then, we would need to consider what it would mean for the law to similarly exhibit non-determinism. 
And this is where ``code is law'' immediately starts to break down. In the ideal case, the law should have deterministic outcomes based on its inputs. 
It can exhibit variation in the relationships between inputs and outputs, but it should not be the case that there is randomness or arbitrariness in those relationships~\cite[pp. 665-666]{kolber2014smoothbumpy}. 
In practice, non-determinism can of course occur in the law. Judicial discretion is not mechanical; given similar inputs, outputs can vary across judges (or even within the same judge)~\cite[p. 78]{tamanaha2004law}. 
But in spite of this non-determinism, sometimes described in relation to the ``indeterminacy thesis,'' the law remains largely predictable.\footnote{As Tamanaha~\cite{tamanaha2004law} discusses, even if there is a relatively small number of unpredictable cases, these cases are often high-impact. 
    General predictability in terms of case numbers should not be misconstrued as a claim that unpredictable cases have low impact. 
    Indeterminacy and unpredictability are more frequent within the Supreme Court, and there always remains the possibility that judges could exploit ``latent indeterminacy'' to suit personal objectives~\cite[pp. 90, 122-3]{tamanaha2004law}.
}  
Contradictions in legal rules, which interfere with predictability, are classically conceived of as ``miscarriages'' of the law~\cite[pp. 38-39]{fuller1965law}. 
Further, as Tamanaha~\cite{tamanaha2004law} argues, there are generally speaking few contradictions in the law, and ambiguities can be handled through ``reasoned analysis.''\footnote{For the indeterminacy thesis ``To have bite it must be shown that existing legal rules form a pervasive mess of contradictions, which critical theorists have not demonstrated''~\cite[p. 88]{tamanaha2004law}.} 

In contrast, non-determinism --- particularly non-stochastic non-determinism --- does not share these qualities. 
Stochasticity perhaps can be considered predictable, its effects reasoned about ``at a higher level of abstraction''~\cite[pp. 3-4]{creel2022leviathan} using probability theory. 
However, from the view of patterns over possible ML outcomes, other forms of non-determinism, ironically, inject unpredictability into ML predictions, both in an intuitive sense and more formally in its resistance to statistical analysis. 
Additionally, as we have seen, empirical work in ML commonly demonstrates that it can result in contradictions with significant consequences~\cite{qian2021variance, forde2021disparate, cooper2021hpo}. 

Moreover, unlike in ML, the legal system embodies answerability. There are actors in the system who must step forward and answer for their decisions; they must provide explanations and are subject to cross-examination~\cite{tribe1971math, brennanmarquez2019plausiblecause}. 
Answerability in part functions to remove randomness and arbitrariness from the law. 
In the long run, the system undergoes an ongoing process of legitimization. 
In other words, the law has mechanisms for recourse, which effectively can serve (however imperfectly) to root out non-determinism; unlike ML, law treats non-determinism vis-à-vis unpredictability and contradictions like a bug, not a feature. 

This indicates a fundamental incompatibility for understanding ML code as law. 
Whereas the law can do work to avoid non-determinism, ML inherently relies on stochasticity and \emph{de facto} relies on non-stochastic non-determinism in state-of-the-art implementations.\footnote{As briefly mentioned earlier, this is a practical reality aimed at optimizing for efficiency under conditions of limited computing resources.} 
The resulting unpredictability of ML code distinguishes it from law in that it causes ML code to evade regulation. 
To borrow a phrase from Jack Balkin, such ``code is lawless''~\cite[p. 52]{balkin2015robotics}; 
the unpredictability that results from non-determinism presents key problems for thinking of code as being constitutive of law.\footnote{Balkin developed this spin on the original refrain in relation to the problem of emergence and unpredictable, unintended consequences in robotic systems. 
    We adopt it more broadly for non-deterministic code.
} 

\section{Conclusion}\label{sec:nondeterminism:conclusion}

Non-deterministic code may itself be lawless, but this does not mean we should entirely avoid its use\footnote{It does, however, seem reasonable to draw the line that ML, if lawless, should not itself be used to design law 
    (e.g., ``Micro-directives [that] will provide \emph{ex ante} behavioral prescriptions finely tailored to every possible scenario''~\cite[p. 2]{casey2015rulesandstandards}). 
    ML can nevertheless still be useful in the service of law, for example, by aiding in the design of tools that help lawyers be more efficient and effective in their work~\cite{delgado2022uncommontask}.
} 
and that we can do nothing to better regulate its deployment in practice. 
On the ML side, we can strive to develop tools that obtain some measure of consistency --- e.g., similar model outcome distributions across training runs --- even in the presence of non-determinism (stochastic or otherwise). 
The current push for more robust ML is in fact working to develop algorithms that leverage non-determinism to learn complex decision surfaces, but also provably have bounded effects on, for example, variance in model-training outcomes. 
In short, ML can do work to tighten distributions/patterns, to provide theoretical limits on error (that then have to be met in practice), and to characterize rigorous trade-offs between computational resource usage for training models and how robust resulting models can be. 
These are rich areas of research in ML, all of which become better-appreciated when understanding ML from a distributional/pattern-level perspective. 

While ML can work improve robustness, stochastic non-de\-ter\-min\-ism will always remain feature, not a bug. 
Legal scholarship thus needs to attend to the role of distributions over outcomes in order to fully appreciate how stochasticity contributes to uncertainty in the behavior of ML systems. A
s we have seen through brief examples concerning unfairness, uncertainty and non-determinism, not just individual outcomes, can themselves implicate harms. 
Since the law will necessarily focus on harms, its work will be to close the gap between these two essential ways of viewing ML --- to ensure that the law is able to reason about distributional aspects in such a way that these aspects serve to clarify how they relate to individual outcomes. 
The law must find ways to bring the distributional and the individual together, such that it can successfully bring ML to account for the harms it causes. 


\begin{acks}
A. Feder Cooper is supported by the Artificial Intelligence Policy
and Practice initiative at Cornell University, the Digital Life Initiative at Cornell Tech, and the John D. and Catherine T. MacArthur
Foundation. The authors would like to thank Joan Feigenbaum, James Grimmelmann, and Solon Barocas for feedback on earlier versions of this work.
\end{acks}

\balance
\bibliographystyle{plainnat}
\bibliography{references}

\end{document}